\documentclass{svproc}
\usepackage{url}

\usepackage[dvipdfmx]{graphicx}        
\usepackage{here}
\begin{document}
\mainmatter              
\title{MEGADOCK-Web-Mito: human mitochondrial protein–protein interaction prediction database}
\titlerunning{MEGADOCK-Web-Mito Database}  
%
\author{Masahito Ohue\inst{1} \and Hiroki Watanabe\inst{2,3} \and Yutaka Akiyama\inst{2}}
\authorrunning{Masahito Ohue et al.} 
%
%
\institute{Department of Computer Science, School of Computing, Tokyo Institute of Technology, G3-56, 4259 Nagatsutacho, Midori-ku, Yokohama City, Kanagawa 226-8503, Japan\\ \email{ohue@c.titech.ac.jp}
\and
Department of Computer Science, School of Computing, Tokyo Institute of Technology, W8-76, 2-12-1 Ookayama, Meguro-ku, Tokyo 152-8550, Japan \email{h\_{}watanabe@bi.c.titech.ac.jp, akiyama@c.titech.ac.jp}
\and
 RWBC-OIL, National Institute of Advanced Industrial Science and Technology, 1-1-1 Umezono, Tsukuba, Ibaraki 305-8560, Japan}

\maketitle              

\begin{abstract}
Mitochondrial diseases are largely caused by dysfunction in mitochondrial proteins. However, annotations of human mitochondrial proteins are scattered across various public databases and individual studies. To facilitate research aimed at elucidating mitochondrial functions, we constructed the MEGADOCK-Web-Mito database as a protein–protein interaction (PPI) prediction data archive, including prediction results for exhaustive protein pairs of 654 mitochondria-related human proteins. MEGADOCK-Web-Mito enables users to search for all PPI prediction results efficiently and comprehensively. In particular, we linked functional annotations to each human mitochondrial protein. The comprehensive and specialized human mitochondrial PPI prediction results and searching function of MEGADOCK-Web-Mito will support further research on mitochondria and mitochondrial diseases.
\end{abstract}
\section{Introduction}
Mitochondria are organelles of eukaryotic cells that serve as the primary energy production sites. Mitochondria are also involved in various other essential cellular processes, including the production of reactive oxygen species, apoptosis, calcium ion storage, and infection defense. In recent years, there has been increased research attention on the range of mitochondrial functions, especially in the context of mitochondrial diseases~\cite{1}. Toward this end, it is essential to elucidate the function of mitochondria-associated proteins.

Many proteins in living organisms interact with each other~\cite{2}, and such protein–protein interactions (PPIs) play a central role in biological phenomena. PPIs themselves, or their dysregulation, have been associated with certain diseases, and drugs have recently been developed to inhibit pathological PPIs~\cite{3,4}.
Along with progress in molecular technologies, enabling the discovery of approximately 196,111 human proteins [based on Swiss-Prot and TrEMBL entries in Universal Protein Resource (UniProt)~\cite{UniProt} as of April 15, 2021], comes the challenge of identifying the large amount of potential PPIs exhaustively in biochemical experiments because of the extensive time and costs involved in such a process. Therefore, computational prediction of PPIs is desirable, which can narrow down the candidate proteins that are most likely to interact with each other prior to experimentation.

To address this challenge, we have developed MEGADOCK~\cite{6,7}, a structure-based PPI prediction method. Although there are various methods for PPI prediction, the structure-based method has the advantage of discovering novel PPIs that do not depend on known interaction information. Moreover, since the three-dimensional structures of approximately 40\%{} of mitochondria-localized proteins have been resolved and reported, structure-based prediction by MEGADOCK is particularly useful for mitochondria-related proteins. We previously reported development of the MEGADOCK-Web database, which is an archive of PPI predictions made by MEGADOCK~\cite{8}. This database implements a search and display function for information on predicted PPIs precomputed by MEGADOCK, and further allows users to browse comprehensive PPI prediction results for approximately 2,000 representative human proteins, including predicted complex structural models.

However, MEGADOCK-Web is a database for general human PPI prediction results, and does not have a specific function specifically related to the mitochondria.
For example, since one of the major causes of mitochondrial diseases is mutation or abnormality of genes with sequence information of mitochondria-related proteins, it is important to be able to refer to mitochondrial-specific gene information.
Therefore, to elucidate the biological functions of mitochondria and the causes of mitochondrial diseases, it is necessary to have a database specialized in mitochondrial PPIs, including information on genes with mitochondrial protein sequence information.

In this paper, we report the results of comprehensive PPI prediction using MEGADOCK for human proteins localized in the mitochondria, including proteins encoded in both the mitochondrial and nuclear genomes. We further introduce our newly developed MEGADOCK-Web-Mito database, which integrates mitochondria-related PPI prediction results and mitochondria-related protein information.

\section{Related Work}
\subsection{MitoProteome}
The MitoProteome is an object-related database developed at the UCSD Supercomputer Center, which contains information on mitochondria-localized proteins that are primarily encoded in the human nuclear genome~\cite{9,10}. Each entry in the MitoProteome corresponds to a gene encoding a protein that is localized in the mitochondria and its basic information, along with annotations of isoforms, splice variants, and functions of the corresponding protein.

\subsection{MEGADOCK-Web}
MEGADOCK-Web~\cite{8} is a PPI prediction results database and browsing system. MEGADOCK-Web contains the results of comprehensive PPI predictions performed by MEGADOCK for representative human proteins.

Fig. 1 shows the structure of the MEGADOCK-Web webpage. The PPI score is a value calculated from the results of PPI prediction calculations by MEGADOCK. Normally, a certain threshold is set for the PPI score, and protein pairs with PPI scores higher than the threshold are predicted to interact.

There are three main functions of MEGADOCK-Web: (1) searching for PPIs predicted by MEGADOCK, (2) visualizing the predicted interaction partners on the Kyoto Encyclopedia of Genes and Genomes (KEGG) pathway map~\cite{KEGG}, and (3) visualizing protein complex models by Molmil viewer~\cite{molmil}.

\begin{figure}[tb]
  \begin{center}
    \includegraphics[width=\textwidth]{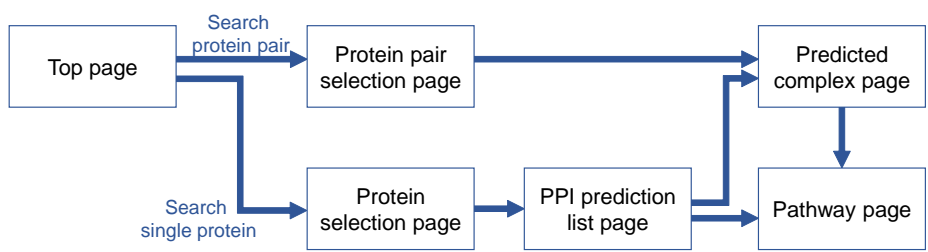}
    \caption{Page transition diagram of MEGADOCK-Web.}
    \label{Fig1}
  \end{center}
\end{figure}

\section{Implementation}
In this study, we developed MEGADOCK-Web-Mito, a database containing predicted results of mitochondria-related PPIs based on the MEGADOCK-Web framework.

\subsection{Data Resources}
The proteins targeted by MEGADOCK-Web-Mito were those that satisfied certain criteria by querying the Protein Data Bank (PDB)~\cite{PDB} via UniProt~\cite{UniProt} among all proteins registered in the MitoProteome. The details of each database are provided below.

\subsubsection{MitoProteome}
We defined mitochondria-associated proteins as those contained in the MitoProteome~\cite{9,10}.
The MitoProteome assigns Mito IDs, represented by ``MT'' and six digits such as ``MT000001,'' to mitochondria-associated protein sequences, and provides a correspondence between the Mito ID and UniProt AC, which is the general ID of the protein sequence.

\subsubsection{PDB}
PDB~\cite{PDB} is a database of the three-dimensional structures of molecules such as proteins and nucleic acids.
Each PDB structure is assigned a four-letter identifier (PDB ID), which may be provided in the form of a complex of multiple molecules such as proteins.
A chain ID was assigned to each molecular chain in the PDB ID structure. For example, the protein structure 1A4I contains two chains, chain A and chain B; therefore, the molecular chain structure is presented as 1A4I\_{}A or 1A4I\_{}B.

\subsubsection{UniProt}
UniProt~\cite{UniProt} is a database that contains protein sequence data and annotation information related to the functions.
UniProt assigns an accession number (UniProt AC) to each protein sequence, and RESTful application programming interfaces (APIs) can be used to retrieve data from UniProt in text, XML, and other formats.

\subsection{Data Collection}
The results of PPI prediction to be included in MEGADOCK-Web-Mito were obtained using the chain structure as input.
Considering that there are multiple possible structures for a given protein, even with an identical sequence, the chain structures of the proteins in the MitoProteome were obtained to satisfy the following conditions:
\begin{enumerate}
    \item The chain structure included in MEGADOCK-Web-Mito is one of the possible structures of the corresponding mitochondria-associated protein.
    \item If there are multiple PDB IDs with the same UniProt AC, MEGADOCK-Web-Mito will include the PDB ID with the highest number of residues resolved. If the number of residues is the same, the structure determined by X-ray crystallography is selected. If there are multiple structures determined by X-ray crystallography, that with the best resolution is selected.
\end{enumerate}
Based on these criteria, a total of 1,675 chain structures were obtained, which were classified into 654 protein types, corresponding to the protein sequence (UniProt AC).

\subsection{PPI Prediction}
The obtained 1,675 chain structures were used for PPI prediction calculations by running MEGADOCK on the TSUBAME 3.0 supercomputer owned by the Tokyo Institute of Technology.
MEGADOCK calculations were performed on $1{,}675 \times 1{,}675 = 2{,}805{,}625$ chain structure pairs.

Suppose the MEGADOCK computation is performed on one f\_{}node with TSUBAME 3.0 [two Intel Xeon E5-2680 V4 processors (14 cores, 2.4 GHz) and four GPUs (NVIDIA TESLA P100 for NVlink-Optimized)], it takes approximately 10 seconds per chain structure pair. An exhaustive PPI prediction calculation using all 1,675 types of chain structures requires approximately 320 node-days. Therefore, we developed a script that uses multiple nodes of TSUBAME 3.0 simultaneously, and achieved exhaustive PPI prediction calculation in only 4.3 days.

\subsection{Link to Mito ID and Entrez Gene ID}
MEGADOCK-Web contains information on proteins, PPI prediction results, and KEGG pathways. 
MEGADOCK-Web-Mito contains this same information, in addition to the Mito ID and Entrez Gene ID. Each ID links to MitoProteome and Entrez Gene~\cite{Entrez} entries to facilitate access to detailed protein information. In addition, a search function according to Mito ID and Entrez Gene ID was implemented.

\subsection{System Architecture}
As for MEGADOCK-Web, we developed MEGADOCK-Web-Mito database using Play Framework~\cite{play}, which is a Java and Scala web application framework that includes components and APIs necessary for web application development. MEGADOCK-Web-Mito adopts a client-server model in the same way as a general web application. Requests from clients are classified into requests to static resources (such as protein PDB files and docking result files) and requests to the web pages. In the former case, Apache directly returns the corresponding resources to clients. In the latter case, Apache redirects requests to the web application running on the local port. We stored information in the relational H2 Database Engine~\cite{h2} database with MySQL~\cite{mysql} wrapped around Play Framework for server-side implementation. Receiving requests from users invokes the web application, which generates HTML dynamically to refer to the database. Server-side implementation is multi-threaded to respond immediately. JavaScript and jQuery are adopted as client-side implementation tools to control the web pages. The web server was built on-premises and can be accessed at the following URL: http://www.bi.cs.titech.ac.jp/megadock-web-mito.

\section{Database Features and Applications}
\subsection{Protein Statistics}
The 654 proteins obtained were classified based on their EC number (enzyme number)~\cite{iubmb}, which are summarized in Table 1.

\begin{table}[tbp]
\begin{center}
\caption{Breakdown of the 654 mitochondria-related proteins in MEGADOCK-Web-Mito.}
\label{tab1}
\begin{tabular}{lr} \hline
class & \#{}protein \\\hline
Oxidoreductases & 97 \\
Transferases & 116 \\
Hydrolases & 94 \\
Lyases & 17 \\
Isomerases & 12 \\
Ligases & 28 \\
Non-enzymes & 290 \\\hline
\end{tabular}
\end{center}
\end{table}

\subsection{Database Structure}
The page structure of MEGADOCK-Web-Mito is shown in Fig. 2, and each page is described in further detail below.

\begin{figure}[tb]
  \begin{center}
    \includegraphics[width=\textwidth]{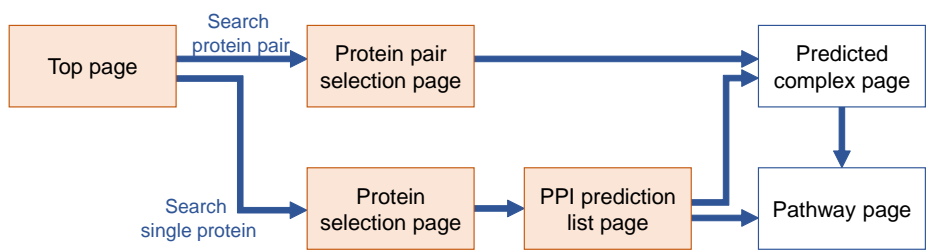}
    \caption{Page transition diagram of MEGADOCK-Web-Mito (pages marked in orange are those that have been modified or extended from MEGADOCK-Web).}
    \label{Fig2}
  \end{center}
\end{figure}

\begin{itemize}
    \item {\bf Top page}\\ The top page of MEGADOCK-Web-Mito has four text fields as search windows, as found in MEGADOCK-Web. As a unique feature of MEGADOCK-Web-Mito, Mito IDs and Entrez Gene IDs have been added to the pull-down menu items placed to the left of each text field, allowing protein searches by Mito ID and Entrez Gene ID, respectively.
    \item {\bf Protein pair selection page and Protein selection page}\\
    In MEGADOCK-Web-Mito, the columns for ``Mito ID'' and ``Entrez Gene ID'' attributes were added to the table of the protein list of search hits. In addition, Mito ID and Entrez Gene IDs were linked to MitoProteome and Entrez Gene, respectively, to enable convenient access to additional information (see Fig.~3).
    \item {\bf PPI prediction list page}\\
     In the PPI prediction list page of MEGADOCK-Web-Mito, we added a row for the ``Mito ID'' and ``Entrez Gene ID'' attributes to the table of protein information and PPI scores (see Fig.~4), which can be searched and linked to the MitoProteome and Entrez Gene ID, respectively. In addition, we added a line that displays a list of KEGG pathways that contain the protein to be searched, enabling the visualization of KEGG pathways for individual proteins. 
\end{itemize}

\subsection{Use Case}
This section describes a specific example of the use of MEGADOCK-Web-Mito to retrieve the PPI prediction information for the protein with Mito ID MT000972 in the MitoProteome, which is related to Parkinson's disease. The following is an example of using MEGADOCK-Web-Mito to retrieve PPI prediction information for the protein with the Mito ID `MT000972.'

\begin{enumerate}
    \item A search for the protein with Mito ID MT000972 in MEGADOCK-Web-Mito revealed the following (Fig.~5): the gene symbol is HTRA2, and there are two chain structures: 2PZD\_{}A and 5FHT\_{}A (Fig.~5(a)). Referring to hsa05012 Parkinson disease from KEGG pathway information (Fig.~5(b)), there is a protein named DJ-1 in the vicinity of HTRA2 (Fig.~5(c)).
    \item Searching for the protein DJ-1 in MEGADOCK-Web-Mito, we found the associated gene symbol PARK7, with Mito ID MT000850. There are four PDB structures (1PDW\_{}A, 1PE0\_{}A, 2RK3\_{}A, 3BWE\_{}A) for this protein.
    \item One of the PPI prediction results for HTRA2 and DJ-1 is shown in Fig.~6, demonstrating the interaction for the structures 5FHT\_{}A and 3BWE\_{}A.
\end{enumerate}

\begin{figure}[H]
  \begin{center}
    \includegraphics[width=\textwidth]{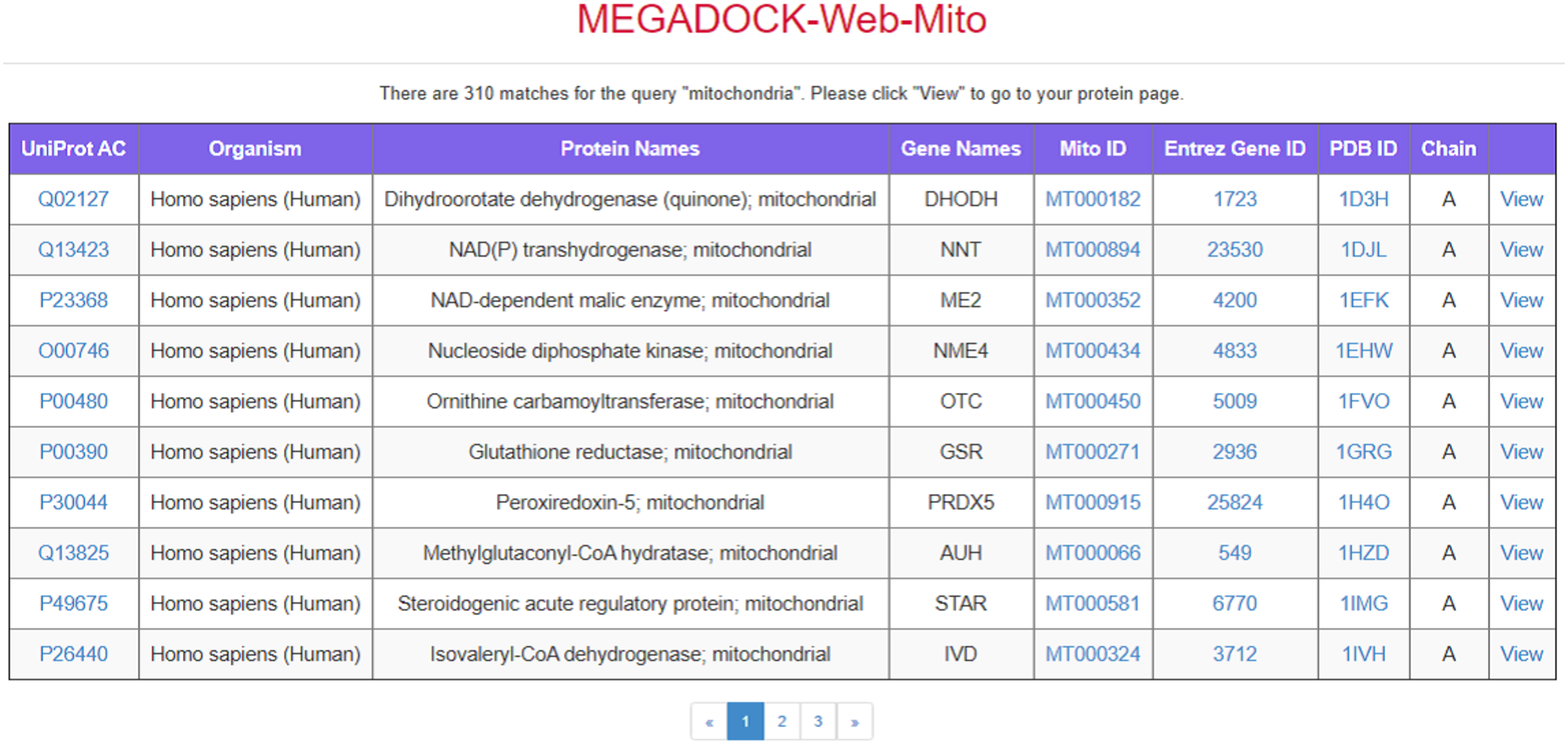}
    \caption{Example of a protein selection page for MEGADOCK-Web-Mito.}
    \label{Fig3}
  \end{center}
\end{figure}

\begin{figure}[H]
  \begin{center}
    \includegraphics[width=\textwidth]{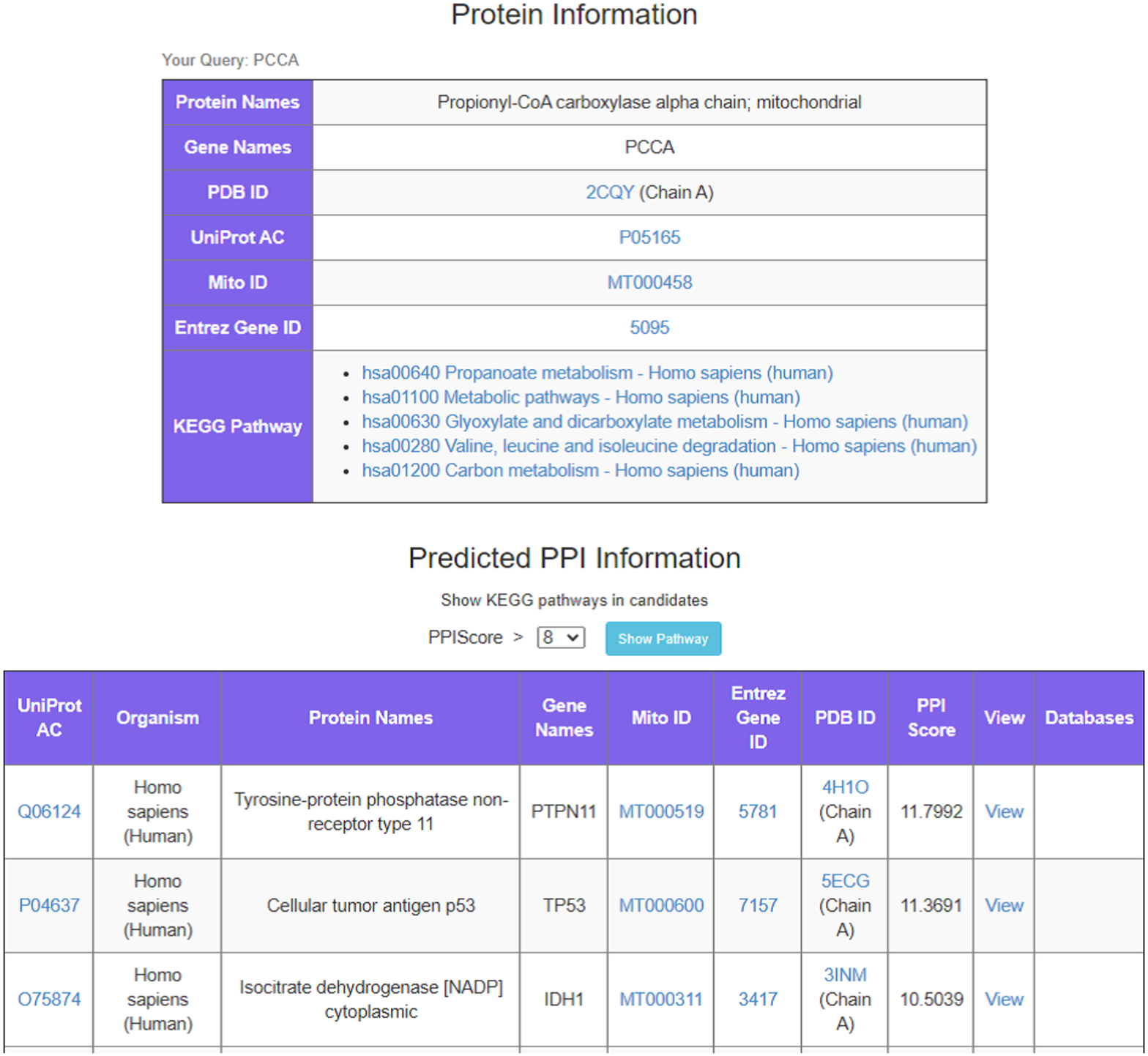}
    \caption{Example PPI prediction list page for MEGADOCK-Web-Mito.}
    \label{Fig4}
  \end{center}
\end{figure}

\begin{figure}[H]
  \begin{center}
    \includegraphics[width=\textwidth]{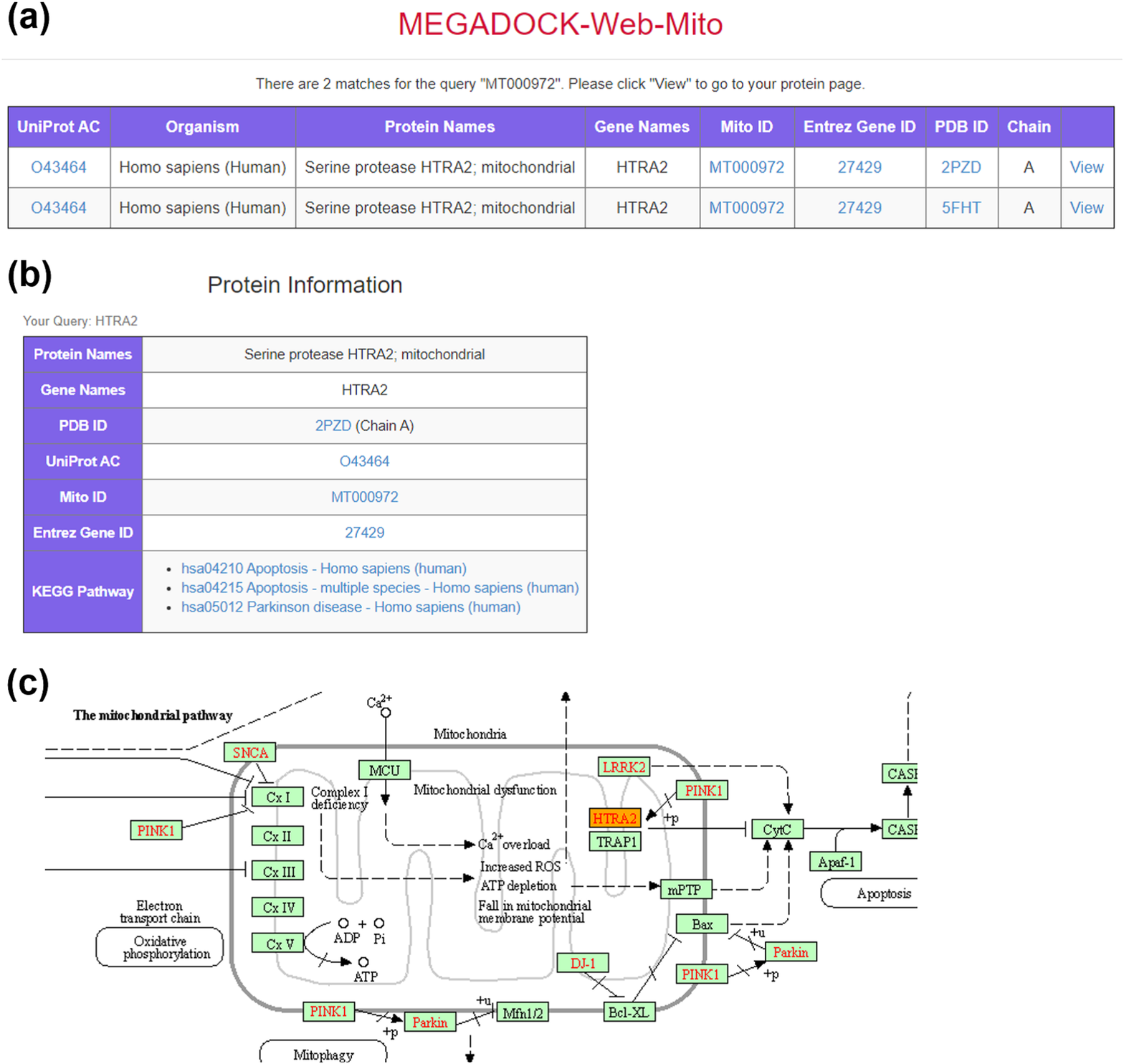}
    \caption{(a) Protein selection page of the result of searching for ``MT000972''. (b) Detailed page of protein with Mito ID MT000972. (c) KEGG pathway map of the Parkinson's disease pathway extracted from the PPI prediction results of HTRA2, and the predicted interaction partners of HTRA2.}
    \label{Fig5}
  \end{center}
\end{figure}
\begin{figure}[H]
  \begin{center}
    \includegraphics[width=\textwidth]{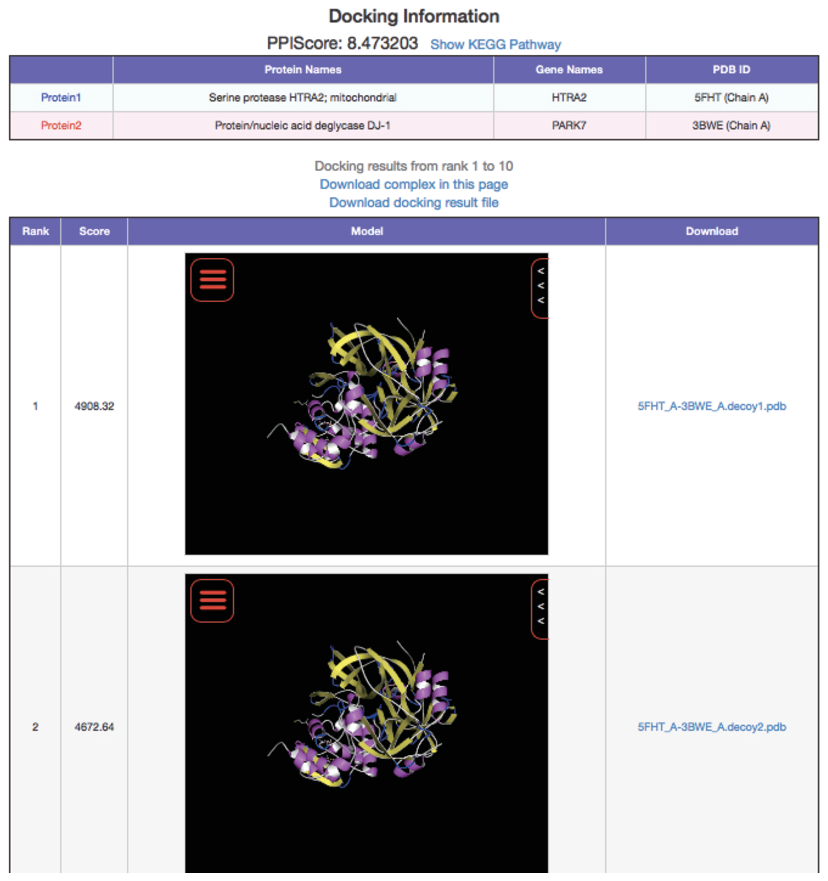}
    \caption{List of predicted complex structures of 5FHT\_{}A and 3BWE\_{}A.}
    \label{Fig6}
  \end{center}
\end{figure}

\section{Conclusions}
In this study, we developed MEGADOCK-Web-Mito, a PPI prediction database related to the mitochondria, based on the existing MEGADOCK-Web framework. 
The PPIs of 654 human mitochondrial proteins with a total of 1,675 protein structures were comprehensively predicted, and the database contains all relevant predictions. In addition, we integrated the annotation information of proteins related to human mitochondria to enable an efficient and detailed investigation of the predicted PPI results.
Since MEGADOCK-Web-Mito has comprehensive PPI prediction results for mitochondria-related proteins and a search function specific to human mitochondria, it is expected to contribute to the study of the molecular mechanisms underlying diseases, particularly mitochondrial diseases.

Although all of the proteins in MEGADOCK-Web-Mito are localized in the human mitochondria, it is also necessary to include proteins localized in the mouse and rat mitochondria, which have similar three-dimensional structures to human proteins, which would further improve PPI prediction results. This will be a challenge for future research.

\section*{Acknowledgements}
This work was partially supported by Japan Society for the Promotion of Science (JSPS) KAKENHI grants (18K18149, 20H04280), Platform Project for Supporting Drug Discovery and Life Science Research [Basis for Supporting Innovative Drug Discovery and Life Science Research (BINDS) grant no. JP20am0101112] from the Japan Agency for Medical Research and Development (AMED). This work was partially conducted as part of the research activities of AIST-Tokyo Tech Real World Big-Data Computation Open Innovation Laboratory (RWBC-OIL). The authors thank Editage (www.editage.com) for English language editing.

%
%

\end{document}